*Review Paper*

# MEMS and ECM Sensor Technologies for Cardiorespiratory Sound Monitoring—A Comprehensive Review

**Yasaman Torabi** [1, *], **Shahram Shirani** [12], *James P. Reilly*[1], *and Gail M Gauvreau*[3]

1. Electrical and Computer Engineering Department, McMaster University, Hamilton, Ontario, Canada
2. L.R. Wilson/Bell Canada Chair in Data Communications, Hamilton, Ontario, Canada
3. Division of Respirology, Department of Medicine, McMaster University, Hamilton, ON L8S 4L7, Canada

* Correspondence: torabiy@mcmaster.ca

**Abstract:** This paper presents a comprehensive review of cardiorespiratory auscultation sensing devices (i.e., stethoscopes), which is useful for understanding the theoretical aspects and practical design notes. In this paper, we first introduce the acoustic properties of the heart and lungs, as well as a brief history of stethoscope evolution. Then, we discuss the basic concept of electret condenser microphones (ECMs) and a stethoscope based on them. Then, we discuss the microelectromechanical systems (MEMSs) technology, particularly focusing on piezoelectric transducer sensors. This paper comprehensively reviews sensing technologies for cardiorespiratory auscultation, emphasizing MEMS-based wearable designs in the past decade. To our knowledge, this is the first paper to summarize ECM and MEMS applications for heart and lung sound analysis.

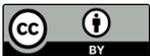

## 1. Introduction

Cardiorespiratory diseases are a leading cause of death all around the world. Therefore, accurate and rapid assessment for signs of such diseases is essential to provide adequate health care for patients [1]. One of the most important signs of cardiovascular disease is cardiac cycle abnormalities. Most respiratory diseases follow abnormality in the respiratory cycle. There are various methods for heart and lung cycle monitoring. Auscultation is one of the most essential diagnosis approaches for the health monitoring of patients.

Human body organs generate acoustic signals that propagate through the tissues and reach the body surface [2]. These acoustic signals are extremely weak but contain a lot of health-related information. To better capture these weak acoustic signals, Rene Laennec invented a mechanical stethoscope [3]. Advancements in technology have revealed that mechanical stethoscopes are not the optimal choice. [4]. Therefore, 3M Corporation invented the electronic stethoscope which converts sound waves into electrical signals through sensors and then amplifies them to improve the results [5].

Current acoustic sensors are mainly divided into two types: electret capacitive sensors and piezoelectric sensors [6]. Electret capacitive sensors are often used in microphones. They operate when a diaphragm and backplate interact with each other after sound enters the microphone and mostly have a junction field effect transistor (JFET) in their input [7]. At their peak of production, condenser microphones became the most widely manufactured type due to their affordability and simple manufacturing process. [8]. However, their signal-to-noise ratio is poor. Additionally, they tend to have limited frequency response and are more susceptible to environmental interference [9]. Therefore, Micro-Electro-Mechanical (MEMS) technology can be utilized for manufacturing sensing structures to address these limitations. MEMS refers to micro-scaled precision devices with mechanical and electronic components [10]. Piezoelectric transducer (PZT) sensors are a type of electroacoustic MEMS sensor that converts the electrical charges produced by solid materials into energy [11].

The market share for MEMS microphones is growing rapidly [12]. This technology minimizes PCB space and lowers the overall manufacturing cost. This highlights the significance of MEMS technology in meeting the demand for bioacoustic devices. Although MEMS microphones are so advantageous, there are still applications where an ECM may be preferred. ECM-based sensors have more circuit design flexibility and are suitable for simple projects. They are also low-cost and accessible.

In this paper, we first introduce cardiac and respiratory cycles, the physiology of the heart and lungs, and their acoustic signals. We then discuss the operational principles of ECM, MEMS, JFET, and Piezoelectricity. Next, we introduce and compare modern sensors for cardiopulmonary auscultation, exploiting ECM and MEMS technology. Finally, we talk about research challenges and prospects in this field.

## 2. Acoustic Properties of Heart and Lung

### 2.1. Cardiac Cycle

The cardiac cycle refers to the synchronized activity of the atria and the ventricles [13]. It is divided into the diastole (heart relaxation) and systole (heart contraction) phases [14]. During the atrial and ventricular diastole, deoxygenated blood enters the right atrium via the superior and inferior vena cava. From there, the blood passes the tricuspid valve and enters the right ventricle. Meanwhile, oxygenated blood flows from the lungs into the left atrium. Then, the blood moves into the left ventricle through the mitral valve [15]. As the ventricular diastole is near to its end, the atrial systole begins, when the atria start contraction. Then, during the ventricular systole, venous blood goes from the right ventricle to the lungs through the pulmonary artery, while arterial blood flows from the left ventricle through the aorta into the circulatory system [16]. The human heart consists of four valves that make the blood flow in only one direction [17]. The atrioventricular valves (mitral and tricuspid valves) separate the atria from the ventricles. The semilunar valves (aortic and pulmonic valves) prevent the blood from flowing back into the ventricles from the aorta and pulmonary arteries (Figure 1a) [18].

The first heart sound (S1) is caused by the closure of the atrioventricular valves and the closure of the aortic valve causes the second heart sound (S2) (Figure 1b). The third and fourth heart sounds (S3 and S4) are two abnormal heart sound components which are the result of early diastole and late diastole respectively [19, 20]. Spencer and Pennington [21] discovered that S1 and S2 appear within the frequency range of 50 to 500 Hz, while S3 and S4 occur between 20 and 200 Hz. The presence of S3 and S4 may suggest heart failure [22]. There are various other heart sounds that may indicate an abnormality. Other abnormal sounds may appear in the heart sounds which are called murmurs. The murmurs can be divided into systolic murmurs, diastolic murmurs and continuous murmurs based on their occurrence location [23]. Rangayyan and Lehner (1987) discovered that murmurs occur at lower frequencies up to 600Hz [24].

## 2.2. Respiratory Cycle

The respiratory cycle consists of two phases: inspiratory and expiratory phases, which are the inhalation of environmental air and exhalation of carbon dioxide, respectively (Figure 1c) [25]. The lungs expand with inspiration and relax during expiration. The muscular diaphragm and the intercostal muscles between the ribs, actively allow the lungs to expand during inspiration [26]. Of all the vital signs including the body temperature, the blood pressure, the cardiac pulse rate, and the respiratory rate, only the respiratory rate can be controlled consciously [27]. Normally, adults breathe 16 to 20 times per minute. Bradypnea occurs when the respiratory rate falls below 16 breaths per minute, while a respiratory rate exceeding 20 breaths per minute is referred to as tachypnea, or rapid breathing [28].

Lung sounds can be divided into three types: breath sounds, vocal resonance sounds, and adventitious sounds [29]. Breath sounds are generated by the lungs during respiration and can be heard across the chest area. The expiratory phase is typically low-pitched, while the inspiratory component is high-pitched and long-lasting [30]. Adventitious sounds are unexpected lung sounds, such as crackles and wheezes [31]. Crackles are discontinuous, explosive sounds caused by the sudden opening and closing of abnormally closed airways [32]. The frequency range of crackles is between 60 Hz and 2 kHz [33]. Wheezes are continuous sounds produced by the constriction of airways resulting from bronchial obstruction, typically within a frequency range of 100 to 1000 Hz [34, 35]. Unlike breath sounds and adventitious sounds, vocal resonance sounds originate in the larynx, not the lungs. In a normal person, speech is incoherent when auscultated over the chest wall due to the filtering effect of lung tissue. However, in the presence of lung consolidation (bronchophony), less attenuation occurs, and voice sounds are heard more clearly over the chest wall [36].

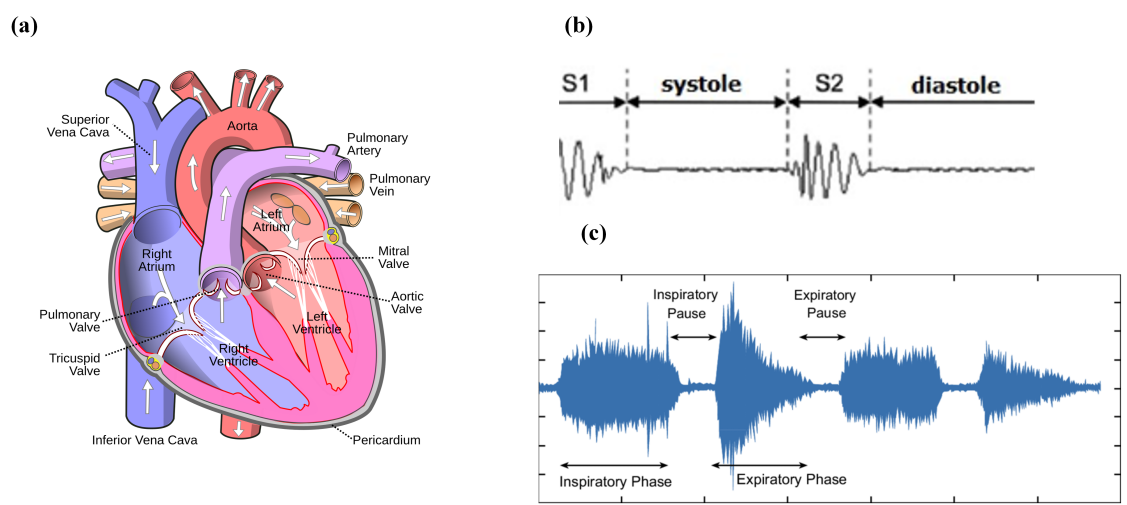

**Figure 1. (a)** The blood flow through the four valves of the heart [37] **(b)** Normal phonocardiogram signal [38] **(c)** The respiratory cycle illustrated via breath sound patterns. The vertical axis is sound intensity, and the horizontal axis is time in seconds [39].

## 3. Evolution of the stethoscope and recent advances

The heart, lungs, and bowels generate subtle acoustic signals that carry valuable diagnostic information [40]. Although these signals are extremely weak, they are crucial for healthcare diagnoses. Recognizing this, Rene Laennec invented the first mechanical stethoscope to amplify these sounds. However, the earliest mechanical stethoscopes encountered a challenge known as the resonance effect. The air cavity within a stethoscope acts as a Helmholtz resonator—a large, enclosed space with a small aperture. This characteristic gives rise to extreme values in their frequency responses at specific frequencies [41] [42]. The resonance effect causes certain frequencies to be amplified or attenuated non-linearly, potentially distorting the captured acoustic signals and complicating accurate diagnosis. Therefore, the digital stethoscope was invented. This amplification method overcomes the high noise and resonance effect [43]. It converts sound waves into electrical signals and then amplifies and processes them.

Current acoustic sensors are primarily categorized into two types: electret capacitive sensors and piezoelectric sensors. Electret capacitive sensors are favored for their low cost and are commonly used in microphone applications. However, they suffer from poor signal-to-noise ratio (SNR) due to considerable energy loss during sound transmission between the body and the transducer [7]. On the other hand, piezoelectric sensors have better SNR thanks to their electromechanical characteristics [45]. These sensors exhibit high sensitivity, rapid response times, and robust mechanical durability, making them suitable for capturing dynamic changes in acoustic signals [46]. Despite these advantages, the similarity between the output signals of piezoelectric sensors and the original acoustic signals is less than 70%, posing challenges in clinical interpretations [47].

Piezoelectric sensors integrated with Microelectromechanical Systems (MEMS) technology can be employed for developing bionic sensing devices. Xue et al. introduced a MEMS underwater microphone inspired by the sensing mechanisms of fish organs [48, 49]. Later, Zhang et al. enhanced the design [50, 51], and Liu et al. proposed a lollipop-shaped prototype with higher sensitivity [52]. Li et al. extended this innovation to applications in heart sound detection [53]. Further advancements were made by Duan and Cui et al., who refined the design into a bat-shaped structure [54, 55, 56]. These developments emphasize the potential of MEMS-based piezoelectric sensors in creating specialized acoustic sensing devices.

## 4. DC- Biased Condenser Microphone

A condenser microphone, as shown in Figure 2, consists of two parallel plates which form a capacitor that linearly converts the distance of its plates into electric voltage, according to equation 1 [57]. In most condenser microphones, a silicon diaphragm is used [58]. When sound waves vibrate the diaphragm, the variable gap between the plates alters capacitance, generating an electrical signal. This signal is then amplified for interpreting its acoustic information.

$$V = q \frac{d}{A\varepsilon_0} \tag{1}$$

where $\varepsilon_0 = 8.8542 \times 10^{12} \ C^2/Nm^2$ is the permittivity constant of free space.

The DC bias is used to power the electronic circuitry of the capacitor and reduce distortion [59]. To achieve high sensitivity, the bias voltage should be as large as possible, but small enough to avoid reducing the dynamic range [60]. This approach allows condenser microphones to achieve low distortion, making them ideal for capturing detailed and accurate sound. However, one disadvantage is that the high bias voltage may lead to a phenomenon known as the pull-in effect, where the moveable membrane may stick to the backplate, causing deformation and failure [61].

## 5. Electret Condenser Microphones (ECM)

The most common alternative to a DC-biased condenser microphone is the electret condenser microphone (ECM), which is shown in Figure 3a. An electret is a dielectric material that has a quasi-permanent electric charge or dipole polarization, which allows it to keep electric polarization over time [62]. While a DC-biased condenser requires an external power supply, the electret condenser uses a pre-polarized diaphragm to generate its own electric field [63]. Electret microphones are mostly made from polymers such as Teflon— PTFE, Teflon—FEP, and Polyvinylidene Fluoride (PVDF) [64-66]. The frequency response of an electret condenser microphone is normally between 20 Hz to 20 kHz [67]. Electret microphones are affordable, small, and have acceptable performance for many general applications. Table 1 summarizes the characteristics of commonly used commercial ECMs.

Table 1. Commercial Electret Condenser Microphones (ECMs)

| Model | Manufacturer | Size (Diameter Ø x Height) (mm) | Frequency Response (Hz) | Sensitivity (dB) | SNR (dB) | Voltage Range (V) |
|---|---|---|---|---|---|---|
| AOM-4544P-2-R [68] | PUI Audio, Inc. | 9.70 Ø× 4.70 mm | 50 Hz ~ 16 kHz | -44dB ±2dB | 60dB | 1.5 ~ 10 V |
| CMA-4544PF-W [71] | CUI Devices | 9.70 Ø× 4.65 mm | 20 Hz ~ 20 kHz | -44dB ±2dB | 60dB | 3 ~ 10 V |
| EM-6050P [72] | Soberton Inc | 6.00 Ø× 5.44 mm | 100 Hz ~ 15 kHz | -42dB ±3dB @ 94dB SPL | 58dB | 1 ~ 10 V |
| RMIC-110-10-6027-NS1 [73] | Raltron Electronics | 6.00 Ø× 2.90 mm | 50 Hz ~ 10 kHz | -42dB ±3dB @ 94dB SPL | 58dB | 1 ~ 10 V |

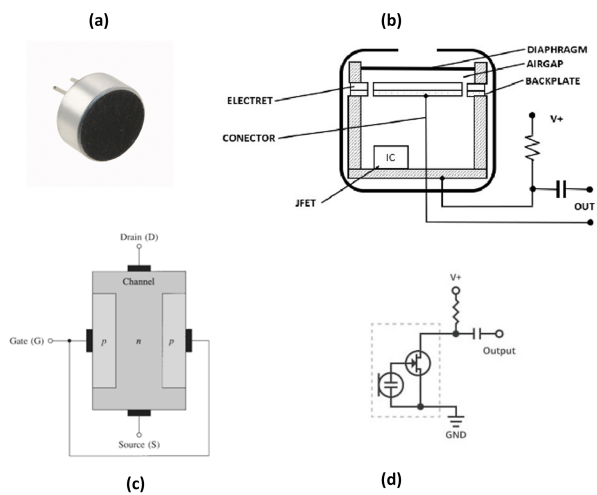

**Figure 3. (a)** Commercial ECM [68] **(b)** The General structure of an electret microphone [65] **(c)** The basic structure of n-channel JFET [69] **(d)** ECM application schematic [70].

An electret microphone consists of a metallized electret diaphragm (with a fixed surface charge) and a back plate (Figure 3b). The air gap acts as the dielectric and forms a variable capacitor. The sound wave moves the diaphragm changing voltage, $\Delta V = \frac{Q}{\Delta C}$. A resistor is positioned between the upper metallization and the back plate. The resistor voltage is amplified and buffered by a junction field effect transistor (JFET). The resulting amplified voltage constitutes the output signal.

It can be shown [60] that for a significantly large resistor $R$ and sine-wave sound, the microphone output voltage is:

$$V = \frac{4\pi\sigma_1 s}{s + s_1} \Delta \sin \omega t \tag{2}$$

where $\sigma_1$ is the electret surface charge, $s$ is the electret thickness, $s_1$ is the air gap, $\omega$ is the circular frequency of the sound wave, and $\Delta$ is the diaphragm displacement.

A JFET is a type of field-effect transistor that uses an electric field to control the flow of current through a semiconductor channel, with terminals known as the source, gate, and drain (Figure 3c). The operation of the device relies on reverse-biasing the PN-junction between the gate and the channel, which controls the channel width and regulates the current flow from the drain to the source. With the PN-junction reverse biased, little current flows into the gate. As the gate voltage (-$V_{GS}$) decreases, the channel narrows until it is "pinched off", stopping the current between the drain and source. At the pinch-off voltage ($V_P$), $V_{GS}$ controls the channel current and $V_{DS}$ has minimal effect. Therefore, the FET acts like a voltage-controlled resistor which has zero resistance. In ECMs, the JFET is typically used in the saturation or "pinch-off" region. This configuration is chosen because in saturation, the JFET operates as a constant current source, which is ideal for signal amplification. [74, 75].

The n-channel JFET characteristics in the saturation (pinch-off) region can be described as follows:

$$I_D = I_{DSS} \left(1 - \frac{V_{GS}}{V_P}\right)^2 (1 + \lambda V_{DS}), \quad (V_P \leq V_{GS} \leq 0, V_{DS} \geq V_{GS} - V_P) \tag{3}$$

Where $\lambda$ is the inverse of the Early voltage and is positive for n-channel devices.

ECMs utilize high-impedance sensors to ensure efficient signal transfer. The JFET's high input impedance, approximately 100 MΩ, makes it an optimal choice for this application [76]. This characteristic also results in a high-pass frequency of around 100 Hz, making the JFET ideal for audio microphones that require high-pass filtering to attenuate low-frequency signals [77].

A key advantage of ECMs is their efficiency in both energy consumption and size. By powering the JFET only, very compact ECMs can be designed. However, it's important to note that ECMs are sensitive to temperature changes, which may restrict their application [78]. The JFET is usually configured in a common-source configuration (Figure 3d), and an external load resistor and DC blocking capacitor are also used in the circuitry [70].

## 6. ECM- based Sensors for Cardiorespiratory Sound Acquisition

While wearable sensors are becoming popular, the large size of traditional cardiac sound probes hinders the design of miniature wearable sensors. To address this challenge,

T. Wang et al. [79] proposed a wearable sound-pressure sensor array which utilizes a field-programmable gate array (FPGA) and a microcontroller unit (MCU), along with a series of denoising methods to enhance accuracy (Figure 4).

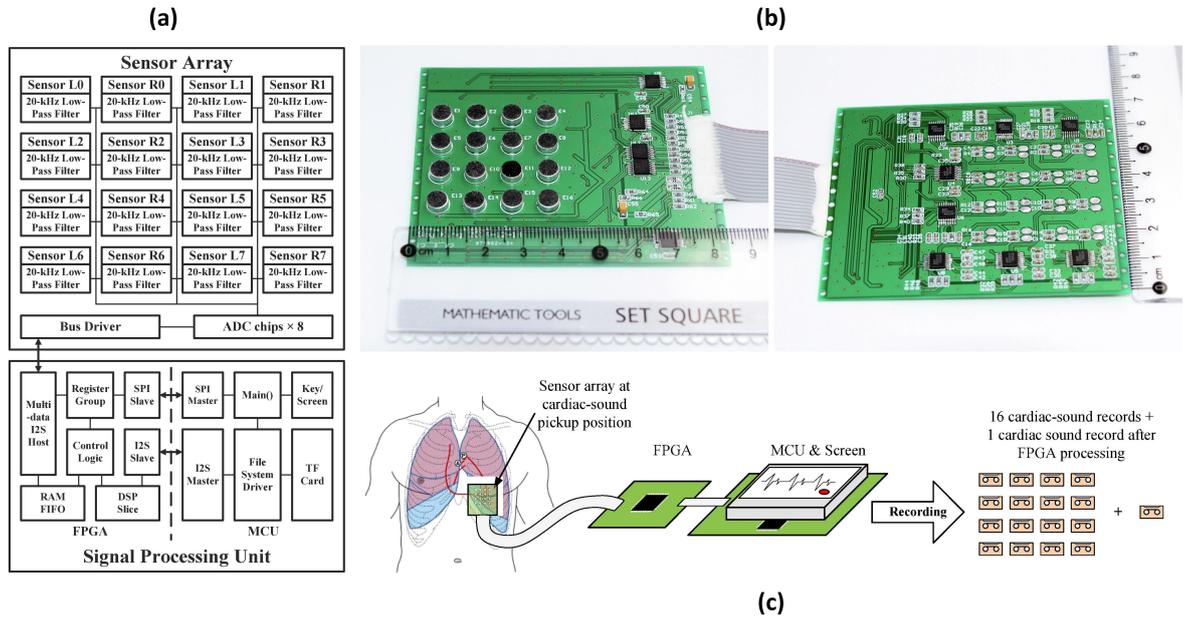

**Figure 4. (a)** Sensor array and signal-processing unit proposed by T. Wang et al. **(b)** Assembled sensor array from front and rear view **(c)** Recording procedure (Images are reprinted with permission from Ref. [79]).

Test results indicated that the noise Root Mean Square (RMS) of the array was more than 3 dB lower than that of a single-sensor structure, demonstrating significant noise reduction. However, further discussion is required regarding several aspects, such as the number of sensors in the array, their arrangement, how the collected data is utilized for diagnostic purposes, and the energy consumption of the device. With the increasing trend towards wearable devices in the wellness monitoring field, the proposed probe has the potential to enable wearable and embedded cardiac-sound monitors if advanced manufacturing techniques such as flexible printed circuits, or self-powered techniques are introduced [80]. Table 2 summarizes the characteristics of the proposed sensor by T. Wang et al. [79].

**Table 1.** Sensor Characteristics

| Parameter | Value |
|---|---|
| Diameter of the ECM sensor | 5 mm |
| Thickness of the ECM sensor | 2 mm |
| Sensor Array Arrangement | 4 × 4 rectangular arrays |
| Dimensions of the Printed Circuit Board | 85 × 70 × 4 mm3 |
| Oversampling Frequency | 48 kHz |
| File Saving Sampling Rate and Format | 8-kHz WAV files |

Further, M. A. A. Hamid et al [81] proposed a low-cost system for recording and monitoring heart sound signals (Figure 5c). The system employs electret microphones (CZN-15E) and amplifiers (NE5534P) to capture and amplify heart sounds. These sounds are then transmitted to a computer via a jack connector, where they are converted from analog to digital signals and de-noised. Experimentally, S1 and S2 peaks were observed, demonstrating the system's effectiveness in capturing detailed cardiac acoustics.

In another study by D. Acosta-Avlos et al. [82], electret microphones were utilized to detect heart sounds in three healthy female participants aged 18 to 29. The sounds were recorded from the thorax, with the microphones securely attached using Micropore® tape and an elastic band. The signals were analyzed using the Fast Fourier Transform (FFT) and autocorrelation functions. The results indicated that the fundamental frequencies of heart signals were similar, although not identical to those detected manually. The study demonstrated that cost-effective electret microphones can achieve a good signal-to-noise ratio, facilitating accurate frequency analysis, although the findings are qualitative due to the small sample size.

Although Fourier Transform methods are widely used in acoustic signal processing, wavelet transform offers a distinct advantage over the short-time Fourier transform (STFT) by providing superior time-frequency resolution. M.V. Shervegar et.al [83] introduced a cost-effective system for heart sound monitoring, featuring an electronic chest piece with a SONY ECM microphone, a pre-amplifier circuit, and a PC or laptop for signal processing. The system employs specific horns for uniform sound capture across a frequency range of 20Hz to 1000Hz and an ultra-low noise LT1115CN8 IC in the pre-amplifier for significant sound amplification (Figure 5d). The recorded PCG signals are stored in .wav format and processed using MATLAB, where filtering and wavelet analysis are applied to identify key heart sound features such as the S1 and S2 peaks (Figure 5a, Figure 5b). This accessible system aims to provide an alternative to expensive digital stethoscopes. The mentioned designs and performance parameters have been presented in Table 3 for a more comprehensive comparison.

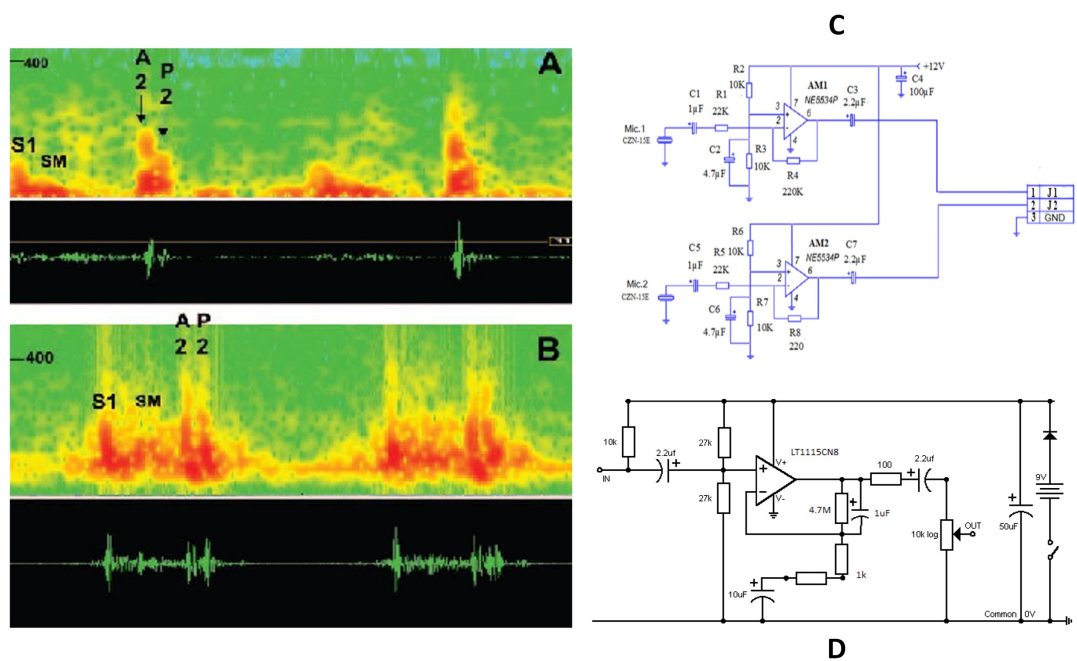

**Figure 5. (a)** Normal heart sound, demonstrating both aortic (A2) and pulmonic (P2) components using the proposed sensor [83] **(b)** Spectral display from a patient. S1 indicates the first heart sound. A systolic ejection murmur (SM) is also displayed in this instance [83] **(c)** The circuit diagrams proposed by M. A. A. Hamid et al [81] **(d)** The circuit diagrams proposed by M.V. Shervegar et.al [83].

Table 3. Summary of methods mentioned above.

| Ref | Technology | Dimension (mm) | Bandwidth (Hz) | Sensitivity (dB) | SNR (dB) |
|---|---|---|---|---|---|
| [79] | Electret Sensor Array FPGA | 5 Ø× 2 mm (Microphone) 85 × 70 × 4 mm³ (PCB) | 1 Hz ~ 1 kHz | N/A | 29.36 dB |
| [81] | CZN-15E ECM Microphone NE5534P Amplifier | 9.7 Ø× 6.7 mm (Microphone) | 20 Hz ~ 16 kHz | -58±2dB (0dB=1V/pa,1kHz) | 60 dB |
| [82] | ECM Microphone Micropore® tape FFT | 3 Ø mm (Microphone) | 1 Hz ~ 10 Hz | N/A | N/A |
| [83] | SONY ECM Microphone LT1115CN8 Amplifier Wavelet | 16.7552 cm³ | 20 Hz ~ 1 kHz | N/A | 90.84 dB |

## 7. Micro-Electro-Mechanical Systems (MEMS)

Although ECM microphones are accurate for general applications, easy to manufacture, and cost-effective, miniature Micro-Electro-Mechanical Systems (MEMS) sensors offer superior sensitivity, lower power consumption, and greater miniaturization, making them ideal for wearable devices [84]. To fabricate these sensors, a MEMS component is mounted on a printed circuit board (PCB). Figure 6a and Figure 6b illustrate the working principle of a capacitive MEMS sensor. MEMS microphones use a diaphragm to form a capacitor that moves in response to sound pressure waves, thereby altering the capacitance to generate an electrical signal. Figure 8c shows a typical MEMS microphone. These microphones can produce either analog or digital outputs. Unlike analog microphones, digital microphones utilize pulse density modulation (PDM) [85, 86], time-division multiplexing (TDM) [87], or the I²S protocol [88, 89] for communication, allowing them to transmit data directly to the processing unit without the need for analog-to-digital converter circuits (Figure 6d, Figure 6e).

The compliance (C) is a parameter used to measure the flexibility of the elastic membrane. The equation describing the electrical sensitivity of the membrane is given by:

$$S_e = A_m C_{eff} \frac{V_{bias}}{g} \tag{4}$$

where $S_e$ denotes the electrical sensitivity, $C_{eff}$ the effective compliance, and $A_m$ the area of the membrane. $V_{bias}$ is the bias voltage of the device and $g$ is the air-gap thickness between the membrane and backplate [91].

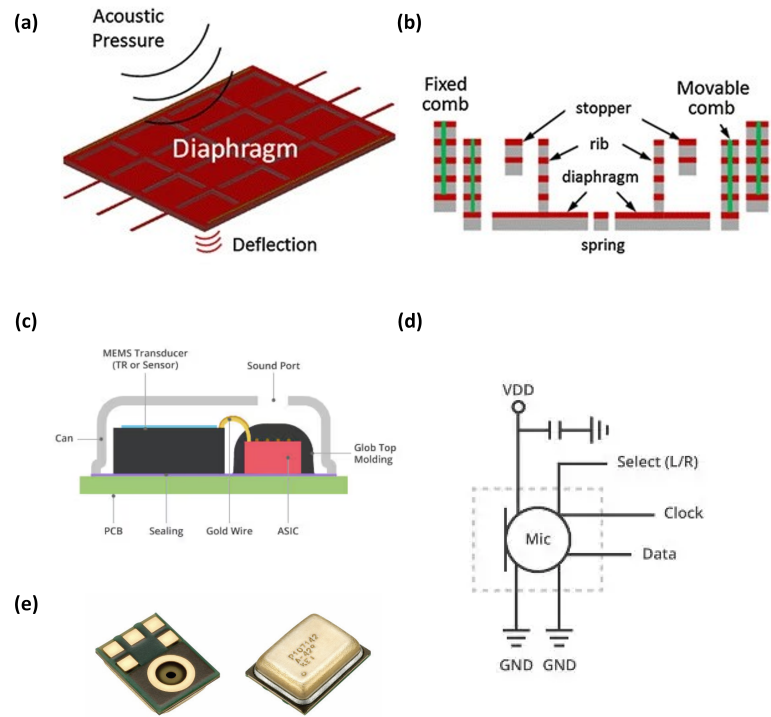

**Figure 6. (a)** MEMS working principle [90] **(b)** MEMS cross-section [90] **(c)** Typical MEMS microphone construction [70] **(d)** digital MEMS microphone application schematic [70].

Piezoelectric transducers (PZT) represent a specialized subset of MEMS technology, providing unique capabilities for acoustic sensing, including but not limited to designing microphones. When force is applied to a piezo material and generates an electric charge, it is known as the piezoelectric effect. Piezoelectric transducers are devices that convert these electrical charges into energy [92]. A piezoelectric microphone features a flexible structure with a back-cavity volume beneath it. Sound pressure deforms this flexible structure, which is at least partially composed of piezoelectric material [93].

This physical structure can be modelled by an equivalent electrical circuit, which facilitates simulations and analyses. It includes an electrical port with a blocked capacitance Ceb and represents acoustic dynamics through Cad and Cab, denoting deformable structure and air compliance, respectively. The transformer ratio Φ illustrates the charge generated across piezoelectric electrodes. The load involves pressure P with impedance Cab, while the rest compromises the transducer system [94].

In 2021, H. Chen et al [95]. introduced a compact and highly sensitive accelerometer designed for continuous monitoring of lung and heart sounds (Figure 7). The device employs a two-stage amplification process, achieving a sensitivity of 9.2 V/g at frequencies below 1000 Hz, thereby making it effective in detecting subtle physiological signals. The sound sensor was constructed using a piezoelectric beam, which featured a top layer made from piezoelectric ceramic materials specifically, lead zirconate titanate (PZT). This beam was coupled with a bottom mechanical layer, with a gap separating the two, and included a movable proof mass made of aluminum (Table 4).

Table 2. Chen et al Sensor Characteristics [95]

|  | Materials | Density (kg/m3) | Young's modulus (GPa) | Size (mm) |
|---|---|---|---|---|
| Piezoelectric beam | PZT | 7.8 × 103 | 66 | 3 × 1 × 0.127 |
| Mechanical beam | Aluminum | 2.7 × 103 | 69 | 3 × 12 × 0.38 |
| Proof mass | Aluminum | 2.7 × 103 | 69 | 20 × 12 × 1.5 |

This advanced sensor successfully identified lung and heart injuries in discharged pneumonia patients for the first time. The SNRs of the lung sound and heart sound signal were 42 dB and 59 dB, respectively.

The sensitivity of the accelerometer can be defined by:

$$\text{sensitivity} = \frac{\text{strain of piezoelectric beam}}{\text{excitation force}} \tag{5}$$

The SNR can be calculated according to the equation below:

$$SNR = 20 \times \log_{10} \frac{Signal\ Voltage}{Noise\ Voltage} \tag{6}$$

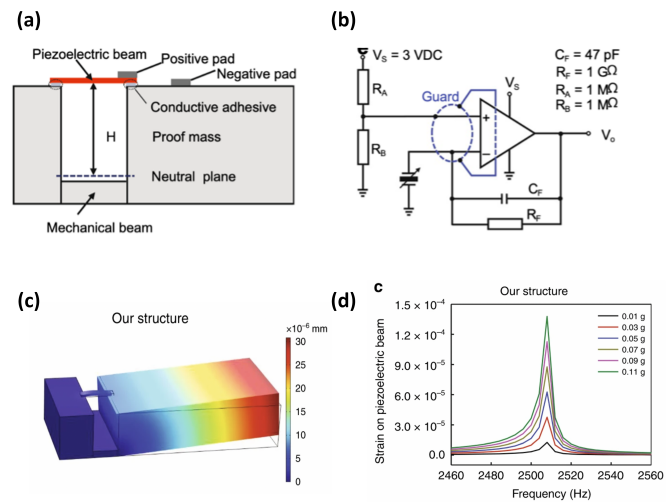

**Figure 7. (a)** Structure of the accelerometer based on an asymmetric gapped cantilever structure **(b)** Built-in charge amplifier circuit for amplification of the piezoelectric signal **(c)** Total displacement of the sensor with proposed structure by Chen et al **(d)** Amplitude-frequency response of the sensor.

The advantages of capacitive MEMS sensors include their small size, mass production capabilities, and low energy consumption, making them suitable for applications that demand high performance and stability, such as heart sound sensors. Table 5 summarizes the characteristics of common commercial MEMS microphones.

Table 5. Commercial MEMS Microphones

| Model | Output Type | Manufacturer | Size (L x W x H) (mm$^3$) | Frequency Response (Hz) | Sensitivity (dB) | SNR (dB) |
|---|---|---|---|---|---|---|
| SPH0645LM4H-B [88] | Digital, I2S | Knowles | 3.50 x 2.65 x 1.10 mm$^3$ | 20 Hz ~ 10 kHz | -26dB ±3dB @ 94dB SPL | 65dB |
| ICS-52000 [87] | Digital, TDM | TDK InvenSense | 4.00 x 3.00 x 1.10 mm$^3$ | 50 Hz ~ 20 kHz | -26 ±1dB @ 94dB SPL | 65dB |
| MM023802-1 [96] | Analog | DB Unlimited | 2.75 x 1.85 x 1.05 mm$^3$ | 30 Hz ~ 10 kHz | -38db ±1dB | 65dB |
| DMM-4026-B-I2S-R [89] | Digital, I2S | PUI Audio, Inc. | 4.00 x 3.00 x 1.10 mm$^3$ | 20 Hz ~ 20 kHz | -26dB ±1dB | 64dB |
| CMM-3526DB-37165-TR [85] | Digital, PDM | CUI Devices | 3.50 x 2.65 x 0.98 mm$^3$ | 100 Hz ~ 10 kHz | -37dB ±1dB @ 94dB SPL | 65dB |
| 3SM121PZB1MB [97] | Analog | 3S (Solid State System) | 4.72 x 3.76 x 1.30 mm$^3$ | 100 Hz ~ 10 kHz | -38dB ±1dB @ 94dB SPL | 68dB |
| IM66D130AXTMA1 [86] | Digital, PDM | Infineon Technologies | 3.50 x 2.65 x 0.99 mm$^3$ | 10 Hz ~ 10 kHz | -36dB ±1dB @ 94dB SPL | 66dB |

**8. MEMS Technology for Cardiorespiratory Acoustic Application**

MEMS technology has revolutionized the analysis of heart and lung sounds. In 2012, Y. Hu et al. [98] developed a chest-worn accelerometer using an asymmetrical gapped cantilever design for continuous cardio-respiratory sound monitoring. This novel design aimed to significantly enhance sensitivity in detecting heart and lung sounds. The MEMS electronic heart sound sensor is designed to optimize the capture of heart sound vibrations within the specific frequency range of heart sounds.

Later in 2016, Zhang et al. [99] noticed a trade-off between the sensor's sensitivity and its first-order resonant frequency. Therefore, they developed a MEMS piezoresistive electronic heart sound sensor featuring a double-beam-block configuration optimized through theoretical analysis and simulations. This configuration raises the first-order resonant frequency while maintaining high sensitivity through stress-concentration grooves. Further, Wang et al. [100] developed a bat-shaped MEMS electronic stethoscope using a novel microstructure design and Wheatstone bridge configuration with piezo-resistors. The Wheatstone bridge design significantly improves the precision of measurements by sensitively detecting minimal resistance variations [101]. In other work, Yilmaz et al. [102] developed a wearable stethoscope integrated into a garment for long-term ambulatory respiratory monitoring. Their design utilized a diaphragm-less transducer, combining silicone rubber and piezoelectric film to capture thoracic sounds effectively. The study highlighted the potential for continuous monitoring of respiratory conditions in real-life settings.

In 2021, Gupta et al. [103] developed a wearable sensor integrating a high-precision accelerometer contact microphone (ACM) to monitor lung sounds. The ACM utilized a nano-gap transduction mechanism, enabling sensitive detection of high-frequency lung sound vibrations. The fabrication process involved advanced lithography to create nano-gap structures, followed by integration into a flexible substrate. The sensor demonstrated high sensitivity and specificity in detecting pathological lung sounds. Later that year, Li et al. [104] utilized a different sensing mechanism and developed a magnetic-induction electronic stethoscope, reducing interference from environmental noise. The resonant frequency of the microstructure can be expressed as Equation 7.

$$f = \frac{1}{2\pi}\sqrt{\frac{Ebh^3}{12m_2L^3}}\sqrt{\frac{6L^2 + 12L + 8}{2L^4 + 7L^3 + 10.5L^2 + 8L + \frac{8}{3}}} \tag{7}$$

where $L$ is the length of the cantilever, $b$ is the width of the cantilever, $h$ is the thickness of the cantilever, $m_2$ is the mass of the induction magnet at the free end and $E$ Is Young's modulus.

This design enhanced sensitivity to low-frequency heart sounds by optimizing the microstructure using low pressure chemical vapor deposition (LPCVD) and Deep Reactive Ion etching of Silicon (DRIE). The process began with the thermal oxidation of a silicon-on-insulator (SOI) wafer to define nano-scale gaps, followed by DRIE etching to create the sensor's cantilever structure. LPCVD was utilized for depositing silicon dioxide to fill the gaps and form the sensing elements. The final device integration included bonding a capping wafer with through-silicon vias (TSVs) for electrical connections. This approach significantly improved the device's performance in noisy environments, making it highly effective for clinical applications in cardiology.

In 2022, Y. Yang et al. [105] and B. Wang et al. [106] focused on enhancing MEMS heart sound sensors through concave designs. Yang et al. proposed an integrated concave cilium structure. Their design featured a bionic-inspired microstructure with high-precision cantilever beams, addressing issues like heart sound distortion and faint murmurs (Figure 8a, Figure 8b). Meanwhile, Wang et al. developed an integrated hollow concave MEMS sensor, emphasizing reduced ciliated mass and expanded bandwidth capabilities over planar designs. Their approach integrated 3D printing for cilium fabrication and MEMS processes for sensor enhancement, highlighting the benefits of concave geometries in enhancing acoustic wave reception for clinical applications. A higher resonant frequency is advantageous because it indicates a greater ability to withstand dynamic loads and vibrations, leading to improved stability and performance in various applications. COMSOL simulation results showed that hollow concave structures have higher resonant frequency than planar ones, enhancing overall functionality (Figure 8c, Figure 8d).

Lastly, current research trends are focusing on enhancing capabilities for real-time and remote monitoring of heart and lung sounds. In S. Hoon Lee et al. [107], S. Hyun Lee et al. [108], and B. Baraeinejad et al. [109] (2023), AI is used for automated diagnostics, demonstrating promising future applications. The first paper detailed a soft, wearable stethoscope employing nanomaterial printing of silicone elastomers and conductive hydrogels for flexible, skin-conformal integration, and utilized convolutional neural networks (CNNs) for real-time cardiopulmonary sound classification, enabling continuous disease diagnosis with high accuracy and effective noise suppression through wavelet denoising (Figure 9a-9d).

The second paper described a flexible lung sound monitoring patch (LSMP) that uses an AI-based breath sound counter with machine learning algorithms for wheeze detection, employing decision trees and support vector machines (SVMs) to classify respiratory sounds in real-time, significantly enhancing long-term respiratory monitoring (Figure 9e, Figure 9f). The third paper introduced a multifunctional digital stethoscope, integrating MEMS microphones and BLE technology, which connects to IoT platforms and employs AI for precise sound analysis and noise management, using digital signal processing (DSP) and machine learning to enhance diagnostic capabilities and remote health monitoring. These innovations highlight advancements in wearable MEMS technologies, emphasizing miniaturization, real-time data processing, and the use of AI for remote patient monitoring and diagnosis.

Table 6 summarizes the characteristics of MEMS designs mentioned above.

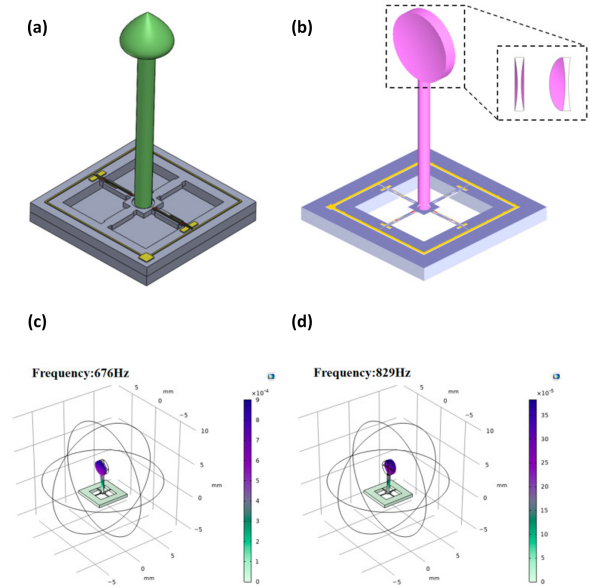

**Figure 8. (a)** Traditional ciliated global microstructural model [105] **(b)** Integrated microstructure model of concave cilium designed by Y. Yang et al. [105] **(c)** Natural frequency in planar microstructures **(d)** Natural frequency in hollow concave micro-structures [106]

**Table 6.** Performance parameters of MEMS designs mentioned above

| Ref | Size (L x W x H) | Bandwidth | Sensitivity | SNR |
|---|---|---|---|---|
| [98] | 35 x 18 x 7.8 mm$^3$ | 20 Hz~ 1kHz | N/A | 65dB |
| [99] | 2 x 2 x 0.02 mm$^3$ | N/A | N/A | 27 dB |
| [100] | 1.2 x 1.2 x 0.02 mm$^3$ | 20 Hz~ 1kHz | −180.7 dB@500 Hz | 38 dB |
| [102] | 37 x 30 x 7.6 mm$^3$ | 100 Hz~ 1.6 kHz | N/A | N/A |
| [103] | 20 x 20 mm$^2$ | ~ 10 kHz | N/A | N/A |
| [104] | 2500 x 12 x 20 μm$^3$ | 20Hz~600 Hz | −189dB@500 Hz | 27.38 dB |
| [105] | 0.1 x 0.34 x 5.7 mm$^3$ | 20Hz~600 Hz | −180.6dB@500Hz | 27.05 dB |
| [106] | 0.12 x 0.34 x 4.9 mm$^3$ | 20Hz~800 Hz | −206.9 dB @200Hz | 26.471dB |
| [107] | 20 x 20 mm$^2$ | 20Hz~1.35 kHz | N/A | 14.8 dB |
| [108] | 40 x 40 mm$^2$ | 100Hz~2 kHz | N/A | N/A |
| [109] | 41 mm diameter | 20Hz~1 kHz | N/A | N/A |

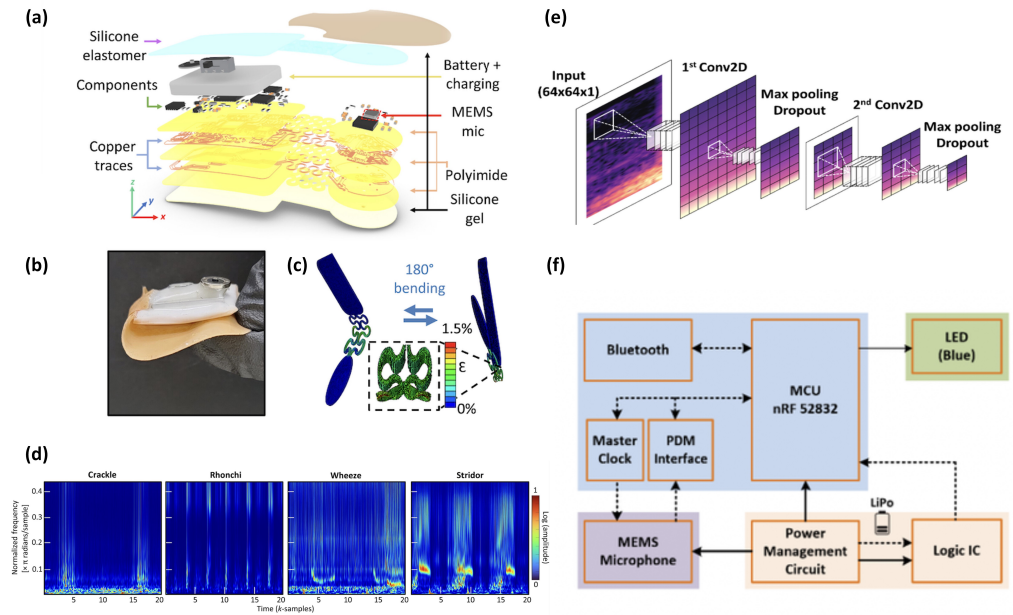

**Figure 9. (a)** Exploded view of S. Hoon Lee's sensor [107] **(b)** Photo of the sensor with 180° bending [107] **(c)** Simulation results showing cyclic bending [107] **(d)** Spectrogram of crackle, rhonchi, wheeze, and stridor data in sample series versus normalized frequency with density for each sample [107] **(e)** Block diagram of the proposed embedded sensor by S. Hyun Lee et al [108] **(f)** Deep learning architecture for training model [108]

## 9. Research challenges and Future perspective

Advancements in electronic stethoscopes are surpassing traditional stethoscopes. Most electronic stethoscopes now feature adjustable filters with various frequency response modes for precise auscultation of heart and lung sounds. Leading manufacturers like 3M, Thinklabs, and Welch Allyn have integrated innovative sensor designs and active noise cancellation techniques to effectively minimize ambient noise, ensuring better diagnostic results [90].

The market share for MEMS microphones is growing rapidly. This newer technology reduces PCB area and final manufacturing cost. Furthermore, semiconductor fabrication technology, alongside the application of audio preamplifiers, enables stable performance characteristics, which is highly beneficial in array structures. Although MEMS microphones have many advantages, there are still applications where ECMs may be preferred. ECM-based sensors offer more circuit design flexibility and are suitable for simple projects. They are also low-cost and accessible.

While MEMS wearables for cardiorespiratory auscultation offer significant advantages, they also face several limitations [110]. First, acoustic signals often have small amplitudes, making them susceptible to ambient noise. Advances in the design of external circuits for active noise cancellation, as well as optimizing the spatiotemporal relationships of sensors in a sensor array are crucial to enhance signal quality. Data acquisition may also present challenges, requiring robust recording positions to account for the variability of cardiorespiratory sound characteristics across different chest locations.

Second, Wearable devices can restrict movement, complicating daily use; collaboration with biomaterials experts could lead to the development of more comfortable fabrics.

Another challenge is the cost of production and recycling, highlighting the need for nanoscale sensing materials with a reduced carbon footprint.

Lastly, as Artificial Intelligence (AI) progresses, auscultation devices require efficient algorithms for automated diagnostics, necessitating collaboration with data scientists and engineers. Current approaches include machine learning methods such as Support Vector Machines (SVM), k-Nearest Neighbor (k-NN), and Neural Networks [111, 112]. Currently, no device offers fully automatic diagnostics, necessitating expert analysis. Future work should consider better automated and real-time platform implementation; most current research relies on local data storage and offline diagnostics. With advances in embedded system design, stethoscopes could become patient-operated. The integration of generative AI technologies, such as Generative Pre-trained Transformers (GPTs) and autoencoders, could further assist in real-time diagnostics and patient interaction, enhancing the overall usability and functionality of these devices.

## 10. Conclusion

This article reviews the progress made in developing auscultation sensing devices for capturing heart and lung sounds over the past decade. To enhance technical depth, it offers a detailed explanation of current research on electronic circuitry, digital signal processing techniques, fabrication processes, and design aspects. To better understand our desired biosensors, we discuss cardiac and respiratory cycles, the physiology of the heart and lungs, the working principles of ECM and MEMS sensors, and related theoretical aspects. This paper discusses several research prototypes and market products from different perspectives, aiding in product selection. It also identifies research challenges and future directions in the field, such as enhancing AI algorithms for embedded sensors to enable real-time monitoring and automatic diagnosis in diagnostic devices.

**References**

[1] Baraeinejad, B. et al., "Design and Implementation of an Ultralow-Power ECG Patch and Smart Cloud-Based Platform," in IEEE Transactions on Instrumentation and Measurement, vol. 71, pp. 1-11, 2022, Art no. 2506811, doi: 10.1109/TIM.2022.3164151.

[2] Cook, J., Umar, M., Khalili, F., Taebi, A., "Body Acoustics for the Non-Invasive Diagnosis of Medical Conditions," Bioengineering (Basel), 2022 Apr 1;9(4):149, doi: 10.3390/bioengineering9040149.

[3] Bishop, P.J. "Evolution of the stethoscope," J. R. Soc. Med., vol. 73, pp. 448–456, 1980, doi: 10.1177/014107688007300611.

[4] Tourtier, J.P. et al. "Auscultation in Flight: Comparison of Conventional and Electronic Stethoscopes," Air Med. J., vol. 30, pp. 158–160, 2011, doi: 10.1016/j.amj.2010.11.009.

[5] Pinto, C. et al. "A comparative study of electronic stethoscopes for cardiac auscultation," Proceedings of the 2017 39th Annual International Conference of the IEEE Engineering in Medicine and Biology Society (EMBC), pp. 2610–2613, 2017, doi: 10.1109/EMBC.2017.8037392.

[6] J. et al. "Design Optimization and Fabrication of High-Sensitivity SOI Pressure Sensors with High Signal-to-Noise Ratios Based on Silicon Nanowire Piezo resistors," Micromachines, pp. 7-187, 2017, doi: 10.3390/mi7100187.

[7] Saveliev, A. et al. "Method of Sensitivity Calculation for Electret Diaphragm Capacitive Sensors," Proceedings of the 2019 12th International Conference on Developments in eSystems Engineering (DeSE); pp. 721–725, 2019, doi: 10.1109/DeSE.2019.00134.

[8] Zawawi, S.A., Hamzah, A.A., Majlis, B.Y., Mohd-Yasin, F., "A Review of MEMS Capacitive Microphones," Micromachines (Basel), 2020 May 8;11(5):484, doi: 10.3390/mi11050484.

[9] Yasuno, Yoshinobu & Miura, Kenzo, (2006). "Sensitivity changes with long-term preservation and practical use of electret condenser microphone," Acoustical Science and Technology, 27, 302-304, doi: 10.1250/ast.27.302.

[10] Chircov, C., Grumezescu, A.M., "Microelectromechanical Systems (MEMS) for Biomedical Applications," Micromachines (Basel), 2022 Jan 22;13(2):164, doi: 10.3390/mi13020164.

[11] Linxian, L. et al. "Package Optimization of the Cilium-Type MEMS Bionic Vector Hydrophone," IEEE Sens. J, vol. 14, pp. 1185–1192, 2014, doi: 10.1109/JSEN.2013.2293669.

[12] MEMS Journal, "MEMS Microphones: Emerging Technology and Application Trends," Date Accessed: 2 Dec 2023. memsjournal.com/2015/07/mems-microphones-emerging-technology-and-application-trends

[13] Tilkian, A.G. et al. "Understanding heart sounds and murmurs: With an introduction to lung sounds," W.B. Saunders, 4th edition, 2001.

[14] Pollock, J.D., Makaryus, A.N., "Physiology, Cardiac Cycle," [Updated 2022 Oct 3]. In: StatPearls [Internet]. Treasure Island (FL): StatPearls Publishing; 2024 Jan-. Available from: ncbi.nlm.nih.gov/books/NBK459327/

[15] Fukuta, H., Little, W.C., "The cardiac cycle and the physiologic basis of left ventricular contraction, ejection, relaxation, and filling," Heart Fail Clin, 2008 Jan;4(1):1-11, doi: 10.1016/j.hfc.2007.10.004.

[16] McDonald, I.G., (1970). "The shape and movements of the human left ventricle during systole: A study by cineangiography and by cineradiography of epicardial markers," The American Journal of Cardiology, 26(3), 221-230, doi: 10.1016/0002-9149(70)90787-3.

[17] Rehman, I., Rehman, A. Anatomy, Thorax, Heart. [Updated 2023 Aug 28]. In: StatPearls [Internet]. Treasure Island (FL): StatPearls Publishing; 2024 Jan-. Available from: ncbi.nlm.nih.gov/books/NBK470256/

[18] Hinton, R.B., Yutzey, K.E., "Heart valve structure and function in development and disease," Annu Rev Physiol, 2011;73:29-46, doi: 10.1146/annurev-physiol-012110-142145.

[19] Ganguly, Antra & Sharma, Manisha, (2017). "Detection of pathological heart murmurs by feature extraction of phonocardiogram signals," Journal of Applied and Advanced Research, 2, doi: 10.21839/jaar.2017.v2i4.94.

[20] Tseng, Y.L., Ko, P.Y., Jaw, F.S., "Detection of the third and fourth heart sounds using Hilbert-Huang transform," Biomed Eng Online, 2012 Feb 14;11:8, doi: 10.1186/1475-925X-11-8.

[21] Spencer, C.S., Pennington, K., "Nurses with Undiagnosed Hearing Loss: Implications for Practice," Online J Issues Nurs, 2015 Jan 5;20(1):6, PMID: 26824264.

[22] Collins, S.P., Lindsell, C.J., Peacock, W.F., & Storrow, A.B., (2004). "Prevalence of S3 and S4 in emergency department patients with decompensated heart failure," Annals of Emergency Medicine, 44(4), Supplement, S98-S99, doi: 10.1016/j.annemergmed.2004.07.320.

[23] Randhawa, S.K., Singh, M., (2015). "Classification of Heart Sound Signals Using Multi-modal Features," Procedia Computer Science, 58, 165-171, ISSN 1877-0509, DOI: 10.1016/j.procs.2015.08.045.

[24] Rangayyan, R.M., Lehner, R.J., "Phonocardiogram signal analysis: A review," Critical Reviews in Biomedical Engineering, 15(3), 211-236.

[25] Li, S., Park, W.H., Borg, A., "Phase-dependent respiratory-motor interactions in reaction time tasks during rhythmic voluntary breathing," Motor Control, Oct 2012;16(4):493-505, doi: 10.1123/mcj.16.4.493.

[26] De Troyer, A., Kirkwood, P.A., Wilson, T.A., "Respiratory Action of the Intercostal Muscles," Physiological Reviews, Apr 2005;85(2), 717-756, doi: 10.1152/physrev.00007.2004.

[27] Registered Nursing Staff Writer. "Respiratory System: TEAS," 2022, Date Accessed: 3 Jan 2023. registerednursing.org/teas/respiratory-system/

[28] Fan, D. et al., "Breathing Rhythm Analysis in Body Centric Networks," IEEE Access, 2018;6, 2846605, doi: 10.1109/ACCESS.2018.2846605.

[29] Dugdale, D.C. "Breath Sounds," 2021, Date Accessed: 3 Jan 2021. mountsinai.org/health-library/symptoms/breath-sounds

[30] Zimmerman, B., Williams, D., "Lung Sounds," [Updated 2023 Aug 28]. In: StatPearls [Internet]. Treasure Island (FL): StatPearls Publishing; 2024 Jan-. Available from: ncbi.nlm.nih.gov/books/NBK537253/

[31] Lowe, R. "Lung Sounds," Date Accessed: 3 Jan 2023. physio-pedia.com/Lung_Sounds

[32] Marques, A., Oliveira, A., "Breath Sounds," Springer International Publishing; Cham, Switzerland: 2018. Normal Versus Adventitious Respiratory Sounds; pp. 181–206. Chapter 10.

[33] Abbas, A., Fahim, A., "An automated computerized auscultation and diagnostic system for pulmonary diseases," J. Med. Syst., 2010;34:1149–1155, doi: 10.1007/s10916-009-9334-1.

[34] Rocha, B.M., Pessoa, D., Marques, A., Carvalho, P., Paiva, R.P., "Automatic Classification of Adventitious Respiratory Sounds: A (Un)Solved Problem?" Sensors (Basel), Dec 24, 2020;21(1):57, doi: 10.3390/s21010057.

[35] Bohadana, A., Izbicki, G., Kraman, S.S., "Fundamentals of Lung Auscultation," N. Engl. J. Med., 2014;370:744–751, doi: 10.1056/NEJMra1302901.

[36] Sarkar, M., Madabhavi, I., "Vocal resonance: a narrative review," Monaldi Archives for Chest Disease, 2024, doi: 10.4081/monaldi.2024.2911.

[37] Ganesan, P. ,"Characterization of Cardiac Electrogram Signals During Atrial Fibrillation," thesis, 2015.

[38] Arslan, A., Yildiz, O. "Automated Auscultative Diagnosis System for Evaluation of Phonocardiogram Signals Associated with Heart Murmur Diseases," Gazi University Journal of Science, vol. 31, no. 1, pp. 112-124, 2018.

[39] Doyle, D.J. "Acoustical Respiratory Monitoring in the Time Domain," The Open Anesthesia Journal, vol. 13, pp. 144-151, 2019, doi: 10.2174/2589645801913010144.

[40] Alemán-Soler, N.M. et al. "Biometric approach based on physiological human signals," Proceedings of the 2016 3rd International Conference on Signal Processing and Integrated Networks (SPIN), pp. 681–686, 2016, doi: 10.1109/SPIN.2016.7566783.

[41] Nussbaumer, M., Agarwal, A., "Stethoscope acoustics," Journal of Sound and Vibration, vol. 539, 2022, p. 117194, ISSN 0022-460X, doi: 10.1016/j.jsv.2022.117194.

[42] Strutt, J.W., "The theory of the Helmholtz resonator," Proceedings of the Royal Society of London. Series A, 1916, pp. 265–275, doi:10.1098/rspa.1916.0012.

[43] Bohadana, A. et al. "Influence of observer preferences and auscultatory skill on the choice of terms to describe lung sounds: a survey of staff physicians, residents, and medical students," BMJ Open Respiratory Research, vol. 7, article no. e000564, 2020, doi: 10.1136/bmjresp-2020-000564

[44] Lella, K.K., Jagadeesh, M.S. & Alphonse, P.J.A. Artificial intelligence-based framework to identify the abnormalities in the COVID-19 disease and other common respiratory diseases from digital stethoscope data using deep CNN. Health Inf Sci Syst 12, 22 (2024). doi: 10.1007/s13755-024-00283-w

[45] Jung, J. et al. "Review of piezoelectric micromachined ultrasonic transducers and their applications," J. Micromech. Microeng, vol. 27, no. 113001, 2017, doi: 10.1088/1361-6439/aa851b.

[46] Seo, Y., Corona, D., Hall, N.A., "On the theoretical maximum achievable signal-to-noise ratio (SNR) of piezoelectric microphones," Sensors and Actuators A: Physical, vol. 264, pp. 341-346, 2017, doi: 10.1016/j.sna.2017.04.001.

[47] Almeida, V.G. et al. "Piezoelectric probe for pressure waveform estimation in flexible tubes and its application to the cardiovascular system," Sens. Actuators A Phys, vol. 169, pp. 217–226, 2011, doi: 10.1016/j.sna.2011.04.048.

[48] Zhang, W. et al."Vector High-Resolution Marine Turbulence Sensor Based on a MEMS Bionic Cilium-Shaped Structure," IEEE Sens. J., vol. 21, pp. 8741–8750, 2021, doi: 10.1109/JSEN.2020.3046836.

[49] Guan, L. "Advancements in technology and design of NEMS vector hydrophone," Microsyst. Technol, pp. 17:459, 2011, doi: 10.1007/s00542-011-1272-4.

[50] Zhang, G.J. et al.** "Design of a monolithic integrated three-dimensional MEMS bionic vector hydrophone," Microsyst. Technol., vol. 21, pp. 1697–1708, 2015, doi:10.1007/s00542-014-2262-0.

[51] Z. et al. "Improvement of the MEMS bionic vector hydrophone," Microelectron. J., vol. 42, pp. 815–819, 2017, doi: 10.1016/j.mejo.2011.01.002.

[52] Liu, Y. et al. "Lollipop-shaped high-sensitivity Microelectromechanical Systems vector hydrophone based on Parylene encapsulation," J. Appl. Phys., vol. 118, no. 44501, 2017, doi: 10.1063/1.4927333.

[53] Li, H. et al. "Design of a high SNR electronic heart sound sensor based on a MEMS bionic hydrophone," AIP Adv., vol. 9, no. 015005, 2019, doi: 10.1063/1.5062619.

[54] Duan, S. et al. "A Bionic MEMS Electronic Stethoscope with Double-Sided Diaphragm Packaging," IEEE Access, vol. 9, pp. 27122–27129, 2021, doi: 10.1109/ACCESS.2021.3058148.

[55] Cui, J. et al. "Design and optimisation of MEMS heart sound sensor based on bionic structure," Sens. Actuators A Phys., vol. 333, no. 113188, 2021, doi: 10.1016/j.sna.2021.113188.

[56] Zhou, C., Zang, C. et al. "Design of a Novel Medical Acoustic Sensor Based on MEMS Bionic Fish Ear Structure," Micromachines, vol. 13, no. 163, 2022, doi: 10.3390/mi13020163.

[57] Shu, Z., Ke, M., Chen, G., Horng, R., Chang, C., Tsai, J., Lai, C., Chen, J., "Design And Fabrication of Condenser Microphone Using Wafer Transfer And Micro-electroplating Technique," DTIP of MEMS & MOEMS, Apr 9-11, 2008.

[58] Li, X., Lin, R., Kek, H., Miao, J., Zou, Q., "Sensitivity-improved silicon condenser microphone with a novel single deeply corrugated diaphragm," Sensors and Actuators A: Physical, vol. 92, pp. 257-262, 2001, doi:10.1016/S0924-4247(01)00582-9.

[59] Printezis, G., Aage, N., Lucklum, F., "A non-dimensional time-domain lumped model for externally DC biased capacitive microphones with two electrodes," Applied Acoustics, vol. 216, 2024, 109758, doi:10.1016/j.apacoust.2023.109758.

[60] Fraden, J. "Handbook of Modern Sensors," SpringerVerlag, 2003.

[61] Lardies, J., Arbey, O., Berthillier, M., "Analysis of the pull-in voltage in capacitive mechanical sensors," International Conference on Multidisciplinary Design Optimization and Applications, Paris, France, 2010.

[62] Gerhard, R., "Dielectric materials for electro-active (electret) and/or electro-passive (insulation) applications," 2019 2nd International Conference on Electrical Materials and Power Equipment (ICEMPE), Guangzhou, China, 2019, pp. 91-96, doi: 10.1109/ICEMPE.2019.8727276.

[63] Audio-Technica, "What are the differences between the microphones that Audio-Technica offers? (Part 3)," Date Accessed: 2 Jan 2024. audio-technica.com/en-us/support

[64] Turnhout, J., "The use of polymers for electrets," Journal of Electrostatics, 1, 147–163, 1975, doi: 10.1016/0304-3886(75)90045-5.

[65] Pyra, Józef & Kłaczyński, Maciej, "Issues of Data Acquisition and Interpretation of Paraseismic Measuring Signals Triggered by the Detonation of Explosive Charges," Sensors, 21, 1290, 2021, doi: 10.3390/s21041290.

[66] Xu, J. & Dapino, Marcelo & Gallego-Perez, Daniel & Hansford, Derek, "Microphone based on Polyvinylidene Fluoride (PVDF) micro-pillars and patterned electrodes," Sensors and Actuators A-physical - SENSOR ACTUATOR A-PHYS, 153, 24-32, 2009, doi: 10.1016/j.sna.2009.04.008.

[67] Zuckerwar, A. J., "Acoustical Measurement," in Encyclopedia of Physical Science and Technology (Third Edition), Ed. R. A. Meyers, Academic Press, 2003, pp. 91-115, ISBN 9780122274107, doi: 10.1016/B0-12-227410-5/00008-9.

[68] PUI Audio, Inc., AOM-4544P-2-R Datasheet, accessed on 5 January 2024, Available online: api.puiaudio.com/file/53873b5a-a40b-45c9-87e9-40627beb81b2.pdf

[69] Sedra, A.S., Smith, K.C. "Microelectronic Circuits," Oxford University Press, Ch. 5, 2004.

[70] Arrow Division. "MEMS vs. Electret Condenser: Which Microphone Technology Should You Use?" 2019, Date Accessed: 3 Jan 2023. arrow.com/en/research-and-events/articles/mems-vs-electret-condenser-which-microphone-technology-should-you-use

[71] CUI Devices, CMA-4544PF-W Datasheet, accessed on 5 January 2024, Available online: cuidevices.com/product/resource/cma-4544pf-w.pdf

[72] Soberton Inc., EM-6050P Datasheet, accessed on 5 January 2024, Available online: soberton.com/wp-content/uploads/2019/02/EM-6050P-14-Feb-2019.pdf

[73] Raltron Electronics, RMIC-110-10-6027-NS1 Datasheet, accessed on 5 January 2024, Available online: raltron.com/webproducts/specs/MICROPHONES/RMIC-110-10-6027-NS1.pdf

[74] Electronics Tutorials. "JFET," Date Accessed: 3 Jan 2023. electronics-tutorials.ws/transistor/tran_5.html.

[75] Leach, W.M., Jr. "The JFET," Georgia Institute of Technology, School of Electrical and Computer Engineering, 2008. leachlegacy.ece.gatech.edu/ece3050/notes/jfet/thejfet.pdf.

[76] Van Rhijn, A., "Integrated Circuits for High Performance Electret Microphones," paper presented at the 114th Convention of the Audio Engineering Society, Amsterdam, The Netherlands, March 22–25, 2003.

[77] Texas Instruments, "Integrated Circuits for High Performance Electret Microphones," Retrieved from ti.com/lit/wp/snaa114/snaa114.pdf, Literature Number: SNAA114, Date Accessed: 2 Jan 2024.

[78] Akerib, D. et al., "Design and performance of a modular low-radioactivity readout system for cryogenic detectors in the CDMS experiment," Nuclear Instruments and Methods in Physics Research A, 5916020, 2008, doi: 10.1016/j.nima.2008.03.103.

[79] Wang, T. et al. "Acoustic-pressure sensor array system for cardiac-sound acquisition," Biomedical Signal Processing and Control, vol. 69, no. 102836, 2021, doi: 10.1016/j.bspc.2021.102836.

[80] Bhatnagar, P. et al., (2023). "Advancing personalized healthcare and entertainment: Progress in energy harvesting materials and techniques of self-powered wearable devices," Progress in Materials Science, 139, 101184, ISSN 0079-6425, doi: 10.1016/j.pmatsci.2023.101184.

[81] Hamid, M. A. A., Abdullah, M., Khan, N. A., AL-Zoom, Y. M. A., "Biotechnical system for recording phonocardiography," International Journal of Advanced Computer Science and Applications, vol. 10, no. 8, pp. 493-497, 2019.

[82] Acosta-Avalos, Daniel & Vitor, Josely & Moraes, Eder, "Detecting the heart and wrist sounds with electret microphones," Academia Letters, 2021, doi: 10.20935/AL677.

[83] Shervegar, M. V., Bhat, G. V., & Shetty, R. M. K., "Phonocardiography – the future of cardiac auscultation," International Journal of Scientific & Engineering Research, 2(10), 1-6, 2011.

[84] Song, P., Ma, Z., Ma, J., Yang, L., Wei, J., Zhao, Y., Zhang, M., Yang, F., Wang, X., "Recent Progress of Miniature MEMS Pressure Sensors," Micromachines (Basel), 2020 Jan 1;11(1):56, doi: 10.3390/mi11010056, PMID: 31906297; PMCID: PMC7020044.

[85] CUI Devices, CMM-3526DB-37165-TR Datasheet, accessed on 5 January 2024, Available online: cuidevices.com/product/resource/cmm-3526db-37165-tr.pdf

[86] Infineon Technologies, IM66D130AXTMA1 Datasheet, accessed on 5 January 2024, Available online: infineon.com/dgdl/Infineon-IM66D130A-DataSheet-v01_00-EN.pdf

[87] TDK InvenSense, ICS-52000 Datasheet, accessed on 5 January 2024, Available online: invensense.tdk.com/wp-content/uploads/2016/05/DS-000121-ICS-52000-v1.3.pdf

[88] Knowles Corporation, SPH0645LM4H-B Datasheet, accessed on 5 January 2024, Available online: mm.digikey.com/Volume0/opasdata/d220001/medias/docus/908/SPH0645LM4H-B.pdf

[89] PUI Audio, Inc., DMM-4026-B-I2S-R Datasheet, accessed on 5 January 2024, Available online: api.puiaudio.com/file/3d9a865c-fddb-4f42-b619-1b96c3a17c74.pdf

[90] Leng, S. et al. "The electronic stethoscope," BioMed Eng OnLine, vol. 14, no. 66, 2015, doi: 10.1186/s12938-015-0056-y.

[91] Chandramohan, G., "Electrical Characterization of MEMS Microphones," tudelft thesis, 2010.

[92] Arnau, A., Soares, D. "Fundamentals of Piezoelectricity," Piezoelectric Transducers and Applications, Springer, 2008, doi: 10.1007/978-3-540-77508-9_1.

[93] American piezo company, "what is a transducer," Date Accessed: 3 Jan 2023. americanpiezo.com/piezo-theory/whats-a-transducer

[94] Seo, Y. et al., "On the theoretical maximum achievable signal-to-noise ratio (SNR) of piezoelectric microphones," Sensors and actuators. A, Physical, vol. 264, pp. 341–346, 2017, doi: 10.1016/j.sna.2017.04.001.

[95] Chen, H. et al., "A two-stage amplified PZT sensor for monitoring lung and heart sounds in discharged pneumonia patients," Microsyst Nanoeng, vol. 7, no. 55, 2021, doi: 10.1038/s41378-021-00274-x.

[96] DB Unlimited, MM023802-1 Datasheet, accessed on 5 January 2024, Available online: dbunlimitedco.com/images/product_images/2D-Drawings/MM023802-1.pdf

[97] 3S (Solid State System), 3SM121PZB1MB Datasheet, accessed on 5 January 2024, Available online: mm.digikey.com/Volume0/opasdata/d220001/medias/docus/5854/3SM121PZB1MB.pdf

[98] Hu, Y. and Xu, Y., "An ultra-sensitive wearable accelerometer for continuous heart and lung sound monitoring," 2012 Annual International Conference of the IEEE Engineering in Medicine and Biology Society, San Diego, CA, USA, pp. 694-697, doi: 10.1109/EMBC.2012.6346026.

[99] Zhang, G.; Liu, M.; Guo, N.; Zhang, W., "Design of the MEMS Piezoresistive Electronic Heart Sound Sensor," Sensors, 2016, 16, 1728, doi: 10.3390/s16111728.

[100] Wang, W. et al., "A bat-shape piezoresistor electronic stethoscope based on MEMS technology," Measurement, vol. 147, 2019, Art. no. 106850, doi: 10.1016/j.measurement.2019.106850.

[101] Carly Riley, "Understanding the Use and Function of MEMS Piezoresistive Pressure Sensors," Date Accessed: 16 June 2024. https://meritsensor.com/understanding-the-use-and-function-of-mems-piezoresistive-pressure-sensors/

[102] Yilmaz, G. et al., "A Wearable Stethoscope for Long-Term Ambulatory Respiratory Health Monitoring," Sensors, vol. 20, 2020, article 5124, Available online: pdfs.semanticscholar.org/7959/c9a0ab8caca56b31d16f3d0d74f60e8a0976.pdf

[103] Gupta, P., Wen, H., Di Francesco, L. et al., "Detection of pathological mechano-acoustic signatures using precision accelerometer contact microphones in patients with pulmonary disorders," Sci Rep, vol. 11, 13427, 2021, doi: 10.1038/s41598-021-92666-2

[104] Li, Y. et al., "Design and verification of magnetic-induction electronic stethoscope based on MEMS technology," Sensors and Actuators A: Physical, vol. 331, 2021, p. 112951, doi: 10.1016/j.sna.2021.112951.

[105] Yang, Y., Wang, B., Cui, J., Zhang, G., Wang, R., Zhang, W., He, C., Li, Y., Shi, P., Wang, S., "Design and Realization of MEMS Heart Sound Sensor with Concave, Racket-Shaped Cilium," Biosensors (Basel), 2022 Jul 18;12(7):534, doi: 10.3390/bios12070534, PMID: 35884337; PMCID: PMC9312695.

[106] Wang, B., Shi, P., Yang, Y., Cui, J., Zhang, G., Wang, R., Zhang, W., He, C., Li, Y., Wang, S., "Design and Fabrication of an Integrated Hollow Concave Cilium MEMS Cardiac Sound Sensor," Micromachines (Basel), 2022 Dec 8;13(12):2174, doi: 10.3390/mi13122174, PMID: 36557472; PMCID: PMC9782983.

[107] Lee, S.H. et al., "Fully portable continuous real-time auscultation with a soft wearable stethoscope designed for automated disease diagnosis," Sci. Adv., 8, eabo5867, 2022, doi:10.1126/sciadv.abo5867

[108] Lee, S.H. et al., "A Wearable Stethoscope for Accurate Real-Time Lung Sound Monitoring and Automatic Wheezing Detection Based on an AI Algorithm," Korea Institute of Science and Technology, Ajou University, The University of Texas at Dallas, Kosin University College of Medicine, 2023, doi: 10.21203/rs.3.rs-2844027/v1

[109] Baraeinejad, B. et al., "Clinical IoT in Practice: A Novel Design and Implementation of a Multi-functional Digital Stethoscope for Remote Health Monitoring," Authorea Preprints, 2023.

[110] Mallegni, N. et al. "Sensing Devices for Detecting and Processing Acoustic Signals in Healthcare," Biosensors, vol. 12, no. 835, 2022, doi: 10.3390/bios12100835.

[111] Messner, E. et al., "Crackle and Breathing Phase Detection in Lung Sounds with Deep Bidirectional Gated Recurrent Neural Networks," 40th Annual International Conference of the IEEE Engineering in Medicine and Biology Society (EMBC), 2018, pp. 356-359, doi: 10.1109/EMBC.2018.8512237.

[112] Torabi, Y., Shirani, S., & Reilly, J. P. (2024). Manikin-Recorded Cardiopulmonary Sounds Dataset Using Digital Stethoscope. arXiv preprint arXiv:2410.03280.

**Disclaimer:**